\documentclass[submission, Phys]{SciPost}
\usepackage{comment}
\usepackage{textcomp}
\usepackage{braket}
\usepackage{amssymb}
\DeclareMathOperator{\Tr}{Tr}

\begin{document}

\begin{center}{\Large \textbf{
Quantum Monte Carlo simulations in the trimer basis: first-order transitions and thermal critical points in frustrated trilayer magnets
}}\end{center}

\begin{center}
L. Weber,\textsuperscript{1*}
A. Honecker,\textsuperscript{2}
B. Normand,\textsuperscript{3,4}
P. Corboz,\textsuperscript{5}
F. Mila,\textsuperscript{4}
S. Wessel\textsuperscript{1}
\end{center}

\begin{center}
{\bf 1} Institute for Theoretical Solid State Physics, JARA-FIT, and JARA-HPC, 
RWTH Aachen University, 52056 Aachen, Germany
\\
{\bf 2} Laboratoire de Physique Théorique et Modélisation, CNRS UMR 8089, \\
 CY Cergy Paris Université, 95302 Cergy-Pontoise, France
\\
{\bf 3} Paul  Scherrer  Institute,  CH-5232  Villigen-PSI,  Switzerland
\\
{\bf 4} Institute  of  Physics,  Ecole  Polytechnique  Fédérale  de  Lausanne 
(EPFL),  CH-1015  Lausanne,  Switzerland
\\
{\bf 5} Institute for Theoretical Physics and Delta Institute for Theoretical 
Physics, University of Amsterdam, Science Park 904, 1098 XH Amsterdam, The 
Netherlands

* lweber@physik.rwth-aachen.de
\end{center}

\begin{center}
\today
\end{center}


\section*{Abstract}
{\bf The phase diagrams of highly frustrated quantum spin systems can exhibit 
first-order quantum phase transitions and thermal critical points even in 
the absence of any long-ranged magnetic order. However, all unbiased numerical 
techniques for investigating frustrated quantum magnets face significant 
challenges, and for generic quantum Monte Carlo methods the challenge is 
the sign problem. Here we report on a general quantum Monte Carlo approach with a loop-update scheme that operates in any basis, and we show that, with an appropriate choice of basis, it allows us to study a frustrated model of coupled spin-1/2 trimers: simulations of the trilayer Heisenberg antiferromagnet in the 
spin-trimer basis are sign-problem-free when the intertrimer couplings are 
fully frustrated. This model features a first-order quantum phase transition, 
from which a line of first-order transitions emerges at finite temperatures 
and terminates in a thermal critical point. The trimer unit cell hosts an 
internal degree of freedom that can be controlled to induce an extensive 
entropy jump at the quantum transition, which alters the shape of the 
first-order line. We explore the consequences for the thermal properties 
in the vicinity of the critical point, which include profound changes in 
the lines of maxima defined by the specific heat. Our findings reveal 
trimer quantum magnets as fundamental systems capturing in full the 
complex thermal physics of the strongly frustrated regime.}

\vspace{10pt}
\noindent\rule{\textwidth}{1pt}
\tableofcontents\thispagestyle{fancy}
\noindent\rule{\textwidth}{1pt}
\vspace{10pt}

\section{Introduction}
\label{sec:intro}

Frustrated magnets typically possess a highly degenerate low-energy 
subspace, because no spin configuration can minimize every interaction 
term simultaneously. This degeneracy can be lifted by collective mechanisms 
leading to (quantum) order-by-disorder~\cite{Starykh2015} or to a kaleidoscope 
of exotic disordered phases, including quantum spin liquids~\cite{Balents2010}.
While some of this physics is captured by certain exactly solvable
models~\cite{Rokhsar1988,Kitaev2003,Kitaev2006}, and the perturbative regimes
close to them~\cite{Balents2004}, many realistic models require numerical
methods for their solution. For two-dimensional (2D) systems, these include 
exact diagonalization~\cite{Meisner2006, Lauchli2019, Wietek2020}, the density 
matrix renormalization group \cite{Yan2011, Depenbrock2012, He2017, Wang2018}, 
and recently developed tensor-network approaches~\cite{Corboz2014,Liao2017,
Shi2019,Kshetrimayum2020}; although both are powerful methods, each has 
specific drawbacks. Another approach is the use of quantum Monte Carlo (QMC) 
methods, which using the stochastic series expansion (SSE) with directed loop 
updates are very efficient for quantum spin systems~\cite{Sandvik1992,
Syljuasen2002,Syljuasen2003}. However, in the presence of geometric 
frustration, QMC typically suffers severely from the negative-sign 
problem~\cite{Troyer2005,Hangleiter2020,Hen2021}.

\begin{figure}[t]
	\centering
	\includegraphics{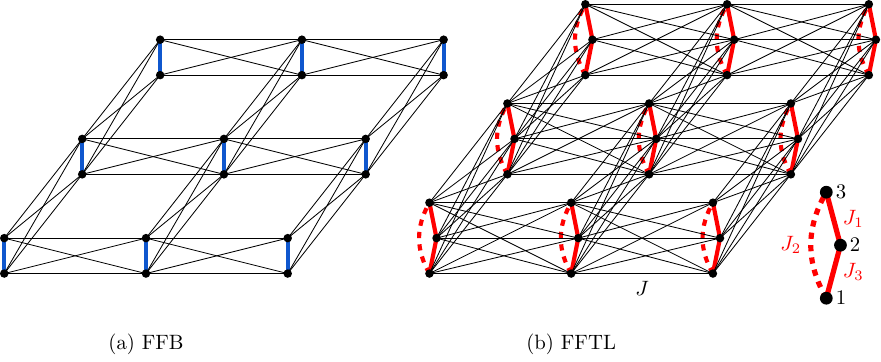}
	\caption{The fully frustrated bilayer (a) and trilayer models (b). In 
the trilayer, the layers are labelled $1$, $2$, $3$ from bottom to top. The 
spins of a trimer unit cell (inset) are coupled by intra-trimer interactions 
$J_1$, $J_2$, and $J_3$ (red) and to all spins in the nearest-neighbor unit 
cells by the same inter-trimer interaction, $J$ (black). The dashed red line 
denoting the coupling ($J_2$) between spins in the bottom (1) and top layers 
(3) of each trimer may differ from the other two couplings ($J_1 = J_3$).}
	\label{fig:ffts_lattice}
\end{figure}

Although the sign problem causes an exponential decrease in the efficiency of 
QMC, it is essentially basis-dependent, and for certain highly frustrated 
models it has been shown that a basis exists in which QMC simulations are 
exactly sign-problem-free (henceforth ``sign-free''). Examples include fully 
frustrated two-leg $S = 1/2$ ladders in one dimension and the fully-frustrated 
bilayer (FFB) [Fig.~\ref{fig:ffts_lattice}(a)] in two~\cite{Honecker2016, 
PhysRevLett.117.197203, Wessel2017, PhysRevB.95.064431, Stapmanns2018}.
In the FFB, the basic components are spin dimers, which are arranged on a 
square lattice and connected symmetrically by nearest-neighbor Heisenberg 
interactions, and the QMC sampling is performed in the local spin-dimer 
basis~\cite{Honecker2016} (related QMC approaches were reported in 
Refs.~\cite{Nakamura98, PhysRevLett.117.197203}). Similar sign-free bases 
exist not only for spin dimers but for arbitrary spin simplices, rendering 
generalizations of these models accessible by QMC~\cite{Honecker2016, 
PhysRevLett.117.197203}. Further, each sign-free model is surrounded by 
an island in parameter space with a tolerable sign problem, which makes QMC 
useful also for understanding related frustrated quantum magnets, such as the 
Shastry-Sutherland lattice (SSL) model~\cite{Wessel2018} by its proximity to 
the FFB. In the SSL model, if the dimer coupling is not too large, then using 
the dimer basis removes the sign problem completely in the zero-temperature 
limit. A similar effect occurs for frustrated quantum magnets in applied
magnetic fields sufficiently far above the saturation field~\cite{DEmidio2020}.

These recent advances in QMC methodology allow us to explore a number of 
important aspects of the thermodynamic properties in specific frustrated 
quantum magnets, and to trace their link to the ground-state properties. 
In the example of the FFB, the spin dimers shape the ground-state phase 
diagram, which is divided into a dimer-singlet quantum disordered and a 
dimer-triplet antiferromagnetic (AFM) phase, separated by a first-order 
quantum phase transition~\cite{MuellerHartmann00}. At finite temperature, it 
was observed that this transition persists, forming a first-order line, 
which terminates at a critical point~\cite{Stapmanns2018}, a scenario that has 
immediate parallels~\cite{Jimenez2020} to the liquid-gas~\cite{Chaikin1995}, 
ferromagnetic~\cite{Kirkpatrick2015}, and Mott transitions~\cite{Georges1996}.

In this work, we study a generalization of the FFB, the spin-1/2 Heisenberg
trilayer with fully frustrated inter-trimer couplings (FFTL), illustrated in 
Fig.~\ref{fig:ffts_lattice}(b). The FFTL constitutes an extension of the 
fully frustrated three-leg ladder~\cite{Honecker2000} to two dimensions, 
similar to the way in which the FFB is an extension of the fully frustrated 
two-leg ladder. In place of the dimers forming the basic components of the 
FFB, the FFTL is composed of trimer unit cells, whose internal frustration 
can be varied from zero at $J_2 = 0$ to maximal at $J_1 = J_2 = J_3$, and we 
will demonstrate that rich physics emerges from this internal degree of 
freedom. Finding this physics in a real material is likely to be a two-step 
process: although the geometry of the FFB has to date been realized only in a 
system with non-Heisenberg spin interactions \cite{Tanaka2014}, almost exactly 
the same physics is found in the compound SrCu$_2$(BO$_3$)$_2$ 
\cite{Jimenez2020}. Similarly, while it is unlikely that a real material would 
possess the precise inter-trimer bonding of Fig.~\ref{fig:ffts_lattice}(b), 
several known quantum magnetic systems are based on triangular clusters with 
frustrated inter-triangle coupling \cite{VanderGriend1999,Millet1999,Luscher2004,
Schnack2004,Aidoudi2011}, and thus are candidates to display some of the phenomenology 
obtained by varying the internal frustration of the FFTL. 

To apply the SSE QMC method to the FFTL model, the sign-free basis is set by 
the local eigenbasis of the trimers. This basis consists of 8 states per unit 
cell, in place of the 4 states in the spin-dimer unit cell, and simulating 
such large local bases usually comes at the cost of significant algorithmic 
complexity. To facilitate both this task and the implementation of specific 
physical operators, we therefore present a formulation of the SSE directed-loop 
update scheme that generalizes readily to arbitrary bases.

With these developments we investigate the thermal physics of the FFTL model 
based on sign-free QMC simulations. In particular, we are able to identify 
a first-order transition line emerging from a first-order quantum phase 
transition, which in the FFTL separates two distinct AFM ground states. 
Similar to the physics of the FFB and SSL~\cite{Stapmanns2018,Jimenez2020}, 
we show that the first-order line of the FFTL is also terminated by a critical 
point, which we identify as belonging to the 2D Ising universality class. This 
phenomenology shares several similarities to the well known phase diagram of 
water, in which the line of first-order transitions in the pressure-temperature 
plane is terminated by a critical point belonging to the 3D Ising universality 
class. However, it was noted in Ref.~\cite{Jimenez2020} that in the 2D quantum 
magnets the specific heat exhibits two pronounced lines of maxima near the 
thermal critical point, in contrast to the single line of specific-heat maxima 
found in water. This behavior can be traced to the special case of the Ising 
model, whose $Z_2$ symmetry in fact enforces an exact symmetry between the 
two specific-heat lines in the magnetic-field-temperature phase 
diagram~\cite{Jimenez2020}. As we report below, varying the intra-trimer 
frustration in the unit cell allows us to (i) control the relative strengths 
of the two lines of maxima in the specific heat of the FFTL, and thus (ii) 
connect the generic ``Ising-like'' behavior observed to date in 2D quantum 
magnets to a more ``water-like'' type of behavior, featuring a single line of 
pronounced specific-heat maxima. Further, we connect the strongest water-like 
behavior, which occurs when the intra-trimer frustration is maximal, to an 
extensive entropy jump across the first-order quantum transition and hence 
an enhanced ``slanting'' of the first-order transition line in the plane of 
``pressure'' (coupling ratio) and temperature. Our observations thus establish 
the FFTL as a generic quantum spin model in which to explore the full 
complexity of the thermal physics uncovered in Refs.~\cite{Jimenez2020, 
Stapmanns2018}, by the unbiased numerical technique of sign-free QMC 
simulations. Taken together, these provide us with the basis on which to 
understand the differences between the superficially distinct forms of 
thermodynamic behavior in certain quantum magnets and in water.  
 
This article is structured as follows. In Sec.~\ref{sec:model}, we provide 
an overview of the FFTL model. In Sec.~\ref{sec:method} we present the 
arbitrary-basis QMC method with which we simulate the FFTL in the trimer basis. 
This allows us in Sec.~\ref{sec:fftl} to analyze the thermal physics of the 
FFTL, discussing in particular the first-order line, the thermal critical 
point, the behavior of the specific heat and of other quantities characterizing 
the critical fluctuations, and the effect of intra-trimer frustration on these 
properties. We conclude in Sec.~\ref{sec:conclusion} by discussing the 
implications of our results and their relation to other recent studies of 
thermal critical points in frustrated quantum magnets.

\section{Model}
\label{sec:model}

In analogy to the well known FFB~\cite{JPSJ-69-878,PhysRevB.66.184402,
PhysRevB.74.144430,LTP07,PhysRevB.82.214412,PhysRevLett.117.197203,
PhysRevB.95.064431,Stapmanns2018} and the fully frustrated three-leg 
ladder~\cite{Honecker2000}, the FFTL is a square lattice of $S = 1/2$ spin 
trimers where all the sites in nearest-neighbor unit cells are connected 
by the same interaction, $J$, giving the Hamiltonian
\begin{align}
H &= \sum_\Delta H_\Delta + H_{\Delta,\Delta+\hat{x}} + H_{\Delta,\Delta+\hat{y}}\nonumber,\\
H_\Delta &= J_1\,\mathbf{S} _{\Delta,2} \cdot \mathbf{S} _{\Delta,3} + J_2\,\mathbf{S} 
_{\Delta,3} \cdot \mathbf{S} _{\Delta,1}+ J_3\, \mathbf{S} _{\Delta,1} \cdot \mathbf{S}
_{\Delta,2}\nonumber,\\
H_{\Delta,\Delta'} &= J \sum_{i,j=1}^3 \mathbf{S}_{\Delta,i} \cdot \mathbf{S}_{\Delta',j}
 = J\,\mathbf{S}_{\Delta} \cdot \mathbf{S}_{\Delta'}.
\end{align}
The sum over $\Delta$ enumerates the trimer unit cells, $\mathbf{S}_{\Delta,i}$ 
is the $i$th spin of trimer $\Delta$, which belongs to the $i$th layer in 
Fig.~\ref{fig:ffts_lattice}(b), and $\mathbf{S}_\Delta = \mathbf{S}_{\Delta,1}
 + \mathbf{S}_{\Delta,2} + \mathbf{S}_{\Delta,3}$ is the total-spin operator of 
trimer $\Delta$. The unit cells are arranged as a square superlattice of 
$L \times L$ trimers.

\begin{figure}
\centering
\includegraphics{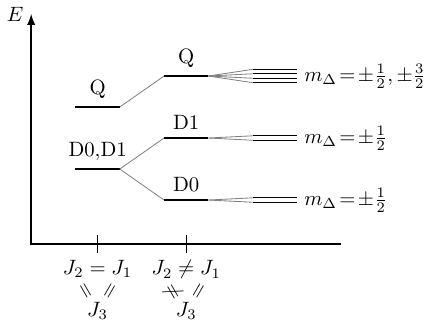}
\caption{Spectrum of a single trimer with spin rotational symmetry. The 8 
states belong to one quartet, Q, and two doublets, D$0$ and D$1$. The doublets 
are degenerate if and only if $J_1 = J_2 = J_3$.}
\label{fig:ffts_spectrum}
\end{figure}

The spectrum of a single trimer consists of 8 states that belong, due to the 
spin rotational symmetry of $H_\Delta$, to one quartet (denoted $\mathrm{Q}$) 
or two doublets ($\mathrm{D}0$ and $\mathrm{D}1$), with respective 
eigenenergies $\varepsilon_\mathrm{Q}$, $\varepsilon_{\mathrm{D}0}$, or 
$\varepsilon_{\mathrm{D}1}$ (Fig.~\ref{fig:ffts_spectrum}). If $J_1 = J_2 = J_3$, 
the two doublets are degenerate, resulting for AFM interactions in a fourfold 
ground-state degeneracy. This is a direct consequence of the symmetry group 
$ G =\mathrm{C}_{3\mathrm{v}}\times$SU(2) at the degeneracy point: the irreducible 
representations of $G$ are tensor products of the irreducible representations 
of the two factors, and one obtains in this case a decomposition of the trimer 
Hilbert space into 
$\mathrm{A}_1\otimes \mathbf{4} \: \oplus  \: \mathrm{E}\otimes\mathbf{2}$.
Away from this point, the doublet levels split and the ground state of a 
single trimer is only doubly degenerate. For $J_1 = J_3$, the eigenstates can 
be classified completely by the expectation values of the commuting operators
\begin{align}
\mathbf{S}^2_\Delta &= l_\Delta (l_\Delta+1),\\
(\mathbf{S}_{\Delta,1}+\mathbf{S}_{\Delta,3})^2 &=l_{\Delta,13}(l_{\Delta,13}+1),\\
S^z_\Delta &= m_\Delta,
\end{align}
where $l_\Delta$ is the total-spin quantum number on trimer $\Delta$ and 
$l_{\Delta,13}$ the total-spin quantum number of spins 1 and 3 on this trimer. 
For convenience we restrict ourselves henceforth to this symmetric case 
(Fig.~\ref{fig:ffts_lattice}).

In the full Hamiltonian, the fully frustrated inter-trimer interaction, $J$,  
couples the total spins, $\mathbf{S}_\Delta$, of adjacent trimers. Because
\begin{align}
    [\mathbf{S}^2_\Delta,S^\alpha_\Delta] &= 0,\\
    \label{eq:comm1}
    [(\mathbf{S}_{\Delta,1}+\mathbf{S}_{\Delta,3})^2,S^\alpha_\Delta] &= 0,\\
    \label{eq:comm2}
    [S^z_{\Delta},S^\alpha_\Delta] &= i \epsilon^{z\alpha \beta} S^\beta_{\Delta} \ne 0,
\end{align}
$l_\Delta$ and $l_{\Delta,13}$ remain good (local) quantum numbers of the 
full model. However, $m_\Delta$ is no longer conserved.

A notable consequence of Eqs.~\eqref{eq:comm1} and \eqref{eq:comm2} is that 
the matrix elements of the fully frustrated inter-trimer interactions in the 
“trimer” basis labelled by $l_\Delta$, $l_{\Delta,13}$, and $m_\Delta$ take the form
\begin{equation}
	\braket{l_\Delta, l_{\Delta,13}, m_\Delta| \mathbf{S}_\Delta |l'_\Delta, 
        l'_{\Delta,13}, m'_\Delta} = \delta_{l_\Delta,l_\Delta'} \delta_{l_{\Delta,13}, 
        l_{\Delta,13}'} \braket{m_\Delta| \mathbf{S}^{S=l_\Delta}_\Delta|m_\Delta'},
\end{equation}
where $\mathbf{S}^{S=l_\Delta}_\Delta$ is a spin-$l_\Delta$ spin operator. The 
matrix elements therefore take the same values in the $S_\Delta^z$ basis as 
in an effective square-lattice AFM of mixed $S = 1/2$ and $S = 3/2$. The 
spin-$S$ Heisenberg AFM on a bipartite lattice can, however, be rendered 
sign-free~\cite{Sandvik2010}, and consequently the inter-trimer Hamiltonian, 
$H_{\Delta,\Delta'}$, of the FFTL in the trimer basis, $\ket{l_\Delta, l_{\Delta,13}, 
m_\Delta}$, is also sign-free. The remaining intra-trimer part, $H_\Delta$, is 
diagonal in this basis and is thus sign-free as well. The trimer basis is 
thus well suited to the construction of a general and efficient SSE 
QMC algorithm, which we present next.

\section{Arbitrary-basis QMC}
\label{sec:method}

The SSE QMC method offers a highly efficient means of simulating quantum spin 
systems. It is based on expressing the quantum Boltzmann operator as an
infinite-order high-temperature expansion with easily evaluated, positive 
coefficients~\cite{Sandvik1992}. This representation, which can be interpreted 
as a classical probability distribution, is then supplemented by the global 
directed-loop~\cite{Syljuasen2002} class of updates that allow an efficient 
sampling of the probability distribution using the Markov-chain Monte Carlo
approach.

In its original and most common formulation, the SSE is carried out explicitly 
in the single-spin $S^z$ basis. However, it is possible to use a different
computational basis, generally at cost of additional algorithmic complexity, 
and alternative bases have been introduced predominantly for one of three 
reasons. First, models with a sign problem in the single-spin basis may be 
“cured” of it in a different basis. This is the case in dimerized Heisenberg 
models with fully frustrated interactions, as introduced in 
Sec.~\ref{sec:intro}, where the enlarged dimer basis gives a fully sign-free 
problem, and the same situation arises in our current trimer-based model.
Second, enlarged bases can also assist calculations for sign-free models 
that nevertheless pose a challenge to the efficiency of SSE QMC updates 
because of high energy barriers, as encountered in the case of Ising-type 
frustration~\cite{Louis2004,Melko2007,Isakov2007,Biswas2016,Emonts2018}. Third, 
changes of basis can make observables 
that are very difficult to compute in one basis readily accessible in another; 
we will meet an example of this in Sec.~\ref{sec:fftl}, where the correlation 
functions of $\mathbf{S}^2_\Delta$ are in essence a four-spin observable.

While formulating the SSE in an alternative basis is straightforward, finding 
a directed-loop Monte Carlo update scheme that works efficiently in a given 
basis often is not. To date such update schemes have typically been implemented 
by hand-crafting loop processes describing the physics of the model to the 
chosen basis; this situation makes the exploration of new bases, for any of 
the three reasons listed above, a somewhat tedious task. 

In this section we present a general scheme to automate this process. We 
start in Subsec.~\ref{sec:opdecomp} by performing the straightforward step 
of arbitrary-basis operator decomposition, which we specialize to the FFTL 
for illustration. In Subsec.~\ref{sec:offdiag} we then detail an “abstract” 
loop-update scheme that can be applied to an arbitrary basis. The performance 
of this scheme is equivalent to a selected, physically motivated class of 
directed-loop updates, but the abstract-update approach generalizes readily 
to arbitrary bases. Particularly for the FFTL, where the trimer basis is 
rather large and the hand-crafted construction of a "proper" loop-update 
scheme (that encodes the full physics) is no longer transparent, the 
abstract-update approach is easy to apply and we will find in 
Sec.~\ref{sec:fftl} that its efficiency is indeed sufficient to perform 
QMC simulations on length scales large enough to address the critical 
properties of the FFTL. 

\subsection{Operator decomposition}
\label{sec:opdecomp}
The starting point for the SSE is the decomposition of the Hamiltonian into 
non-branching operators (those which map each basis vector to a single basis 
vector, and not to a superposition of different basis vectors). We express  
the FFTL model Hamiltonian in the form 
\begin{equation*}
H = \sum_{\braket{\Delta,\Delta'}} \tilde{H}_{\Delta, \Delta'},\quad\tilde{H}_{\Delta,\Delta'}
 = H_{\Delta,\Delta'} + \frac{1}{4} \left(H_\Delta+H_{\Delta'}\right),
\end{equation*}
where the intra-trimer terms have been absorbed into the inter-trimer part,
and decompose $H$ into the operators
\begin{align}
H &= \sum_{\substack{\braket{\Delta,\Delta'}\\x_\Delta, x_{\Delta'}\\ y_{\Delta}, y_{\Delta'}}} 
\braket{x_\Delta, x_{\Delta'}|\tilde{H}_{\Delta,\Delta'}|y_\Delta, y_{\Delta'}} 
\ket{x_\Delta,x_{\Delta'}}\!\bra{y_\Delta,y_{\Delta'}} =: \sum_{\alpha} h_\alpha,
\end{align}
denoting the sets of trimer-basis states $\ket{l_\Delta,l_{\Delta,13},m_\Delta}$ for 
brevity by $\ket{x_\Delta}$ and $\ket{y_\Delta}$. The operators $h_\alpha$ of this 
decomposition are labelled by the bond $\braket{\Delta,\Delta'}$ and for 
notational simplicity we combine the summation indices into a single 
multi-index, $\alpha$. Using this decomposition, the 
partition function can be expanded into a form that is simple to evaluate 
because it contains only sums over non-branching operator strings (details 
may be found in Ref.~\cite{Sandvik2010}),
\begin{align}
	Z = \Tr e^{-\beta H} &= \sum_{n=0}^\infty \sum_{\psi_0} \frac{\beta^n}{n!} 
        \braket{\psi_0| (-H)^n|\psi_0}\\ &= \sum_{n=0}^\infty \sum_{\psi_0} 
        \sum_{\{\alpha_p\}_{p=1}^n} \frac{\beta^n}{n!} \braket{\psi_n|\psi_0} 
        \prod_{p=1}^{n} \braket{\psi_{p-1}|(-h_{\alpha_p})|\psi_p},
\end{align}
where $\ket{\psi} = \ket{x_1,x_2,x_3,\cdots}$ is the trimer-product basis of 
the full Hilbert space and the sum is over all $n$-sequences of $\alpha$. The 
sequence of intermediate states $\{\psi_1,\dots,\psi_n\}$ is generated by the 
successive application of the non-branching operators $h_{\alpha}$ to the trace 
state $\ket{\psi_0}$.

To increase numerical efficiency, the sequences of $h_\alpha$ are in practice 
padded by identity operators so that they always have a fixed length, 
$\mathcal{L}$. This length is chosen adaptively to be sufficiently large 
that it has no measurable effect on the result. However, the padding procedure 
introduces a combinatorial factor that must be accounted for in the padded 
expansion (so that $n$ now denotes the number of non-identity operators 
contained in the operator string $\{\alpha_p\}$), leading to 
\begin{align}
	\tilde{Z} &= \sum_{\psi_0, \{\alpha_p\}_{p=1}^\mathcal{L}} \frac{\beta^n 
        (\mathcal{L}-n)!}{\mathcal{L}!} \braket{\psi_\mathcal{L}|\psi_0} 
        \prod_{p=1}^{\mathcal{L}} \braket{\psi_{p-1}|(-h_{\alpha_p})|\psi_p} =: 
        \sum_{\mathcal{C}} W(\mathcal{C}).
\end{align}
Because the weights $W(\mathcal{C})$ are non-negative for the FFTL in the 
trimer basis, the quantum partition function can be interpreted as a classical 
probability distribution and different observables can be calculated using 
Markov-chain Monte Carlo methods.

\subsection{Abstract directed-loop update}
\label{sec:offdiag}

To sample from a Markov chain, we need an ergodic update for the configuration 
$\mathcal{C} = \{ \psi_0, \{\alpha_p\}\}$ that fulfills the detailed-balance 
criterion. A combination of two types of update is used widely in the 
context of the SSE. First, the diagonal update inserts or removes diagonal 
operators (in the computational basis) locally in the string 
$\{\alpha_p\}$~\cite{Sandvik1992}. Second, for the off-diagonal operators and 
the state $\psi_0$, global loop updates respecting the trace boundary condition 
on $\braket{\psi_\mathcal{L}|\psi_0} \ne 0$ are employed~\cite{Syljuasen2002, 
Alet2005}. In general, directed-loop updates function as follows: starting 
from a given SSE configuration, two discontinuities are introduced at lattice 
site $i$ with operator index $p$. Referring to these discontinuities as 
the loop head and tail, the head is propagated stochastically through the 
SSE operator string until it meets with the tail. At this point the two 
discontinuities can annihilate each other again, leaving a loop along which 
the spin configuration has changed.

The diagonal updates generalize readily to arbitrary bases, and so can be 
applied directly. For the directed-loop updates, however, the situation is 
more complicated. As noted above, approaches to date have tended to involve 
crafting the loops for each specific model by associating the possible 
discontinuities with physically intuitive operators. As an example, the 
operators $S^+$ and $S^-$ are used for spin systems in the conventional 
single-site $S^z$ basis. For local bases containing more than a single 
site, making this identification becomes increasingly difficult, because 
the number of basis states grows faster than the number of intuitive 
operators. An example from the dimer basis is the inclusion of the 
spin-difference operators, $\mathbf{D} = \mathbf{S}_1 - \mathbf{S}_2$, 
in addition to normal spin flips~\cite{Honecker2016,Wessel2017}. For the 
trimer basis, spin-difference operators alone are not sufficient, a result 
we illustrate by expressing in terms of spin components the operator that 
flips both $l_{\Delta,12}$ and $m_\Delta$ simultaneously,
\begin{align}
& \sum_{l_{\Delta,12}, m_\Delta} \ket{1/2, l_{\Delta,12}, m_\Delta}\bra{1/2, 1 - 
l_{\Delta,12}, -m_\Delta} \nonumber \\
& \qquad\qquad\qquad = \frac{4}{\sqrt{3}} \left[S_2^x (S_1^y S_3^y + S_1^z 
S_3^z - 1) - S_1^x (S_2^y S_3^y + S_2^z S_3^z - 1)\right].
\end{align}
While it may be possible to simplify such an expression, and to introduce 
abbreviations for its operators, it is apparent that the language of physical 
loop operators becomes increasingly complex in larger cluster bases.

The alternative route we introduce is an abstract directed-loop update 
scheme that can be constructed canonically for any basis or model. As a 
general starting point, we associate the discontinuities at the loop head 
and tail not with operators but with bijective functions or “actions,”
\begin{equation}
	a: X \rightarrow Y;\quad X, Y \subset \mathcal{B},
\end{equation}
which map the state before the discontinuity, $x$, to the state after the 
discontinuity, $a(x)$, both states being elements of the local (trimer or 
other arbitrary) computational basis, $\mathcal{B}$. In this picture, the 
equivalent of $S^+$ and $S^-$ for the single-spin $S^z$ basis would be the 
two actions $a^{\pm}: \ket{m} \mapsto \ket{m\pm 1}$, where $m$ denotes the 
$S^z$ eigenvalue. Note that the domain $X$ and image $Y$ of an action $a$ 
may not include all of the basis states, as is the case here for $a^+$ and 
the state $\ket{m=S}$.

\def\lin{l_\text{in}}
\def\lout{l_\text{out}}
\def\ain{a_\text{in}}
\def\aout{a_\text{out}}

\begin{figure}
	\includegraphics{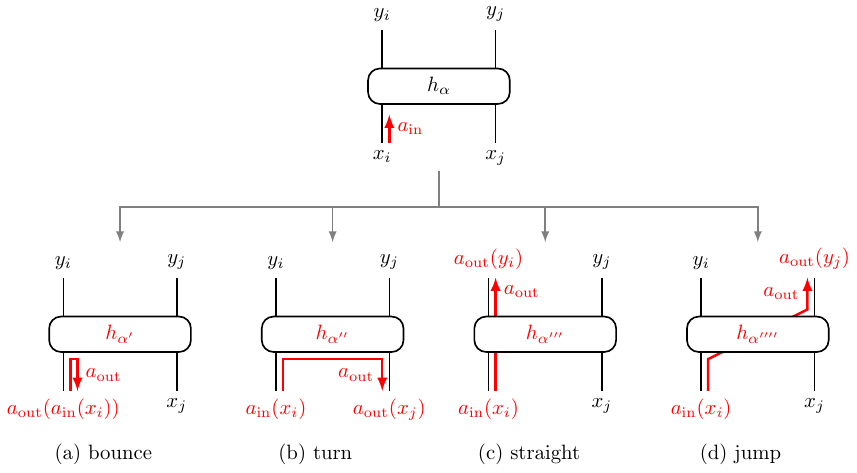}
	\caption{Abstract directed-loop updates. Whenever the loop head 
arrives at leg $\lin$ of an operator $h_\alpha$ carrying the action $\ain$, 
it may continue on a different leg, $\lout$, with a different action, $\aout$. 
Along the path of the head, the four states $x_i$, $x_j$, $y_i$, and $y_j$ 
defining $h_\alpha$ are changed by the application of $\ain$ and $\aout$.}
	\label{fig:opscatter}
\end{figure}

The loop head moves through the configuration carrying one of these actions, 
and whenever it hits an operator $h_\alpha$ there are different ways to proceed, 
as represented in Fig.~\ref{fig:opscatter}. The head can change its real-space 
position and direction by hopping to a different operator leg, $l$, or it can 
change the type of its discontinuity, meaning its action $a$. Depending on 
these choices, the operator is changed, $h_\alpha \rightarrow h_{\alpha'}$, after 
the head has passed. A way to bring this change of configuration into 
agreement with detailed balance is to demand that each junction, or scattering 
event, within the propagation process itself fulfills detailed balance.

Adopting the notation of actions $a$ and legs $l$, each operator scattering 
event can be described as a transition
\begin{equation}
	\alpha,\ain, \lin \rightarrow \alpha',\aout, \lout.
\end{equation}
If the operator before scattering was $h_\alpha = -w_{\alpha} \ket{x_1,x_2} \! 
\bra{x_3,x_4}$, the new operator, $h_{\alpha'}$, is the one where first $x_{\lin}$ 
is replaced by $x'_{\lin} = \ain(x_{\lin})$, and in a second step the state at 
$\lout$ is set to $x''_{\lout} = \aout(x'_{\lout})$.
The detailed-balance condition for this process is then 
\begin{equation}
	\label{eq:detbal}
w_\alpha\,p(\alpha, a_\text{in}, l_\text{in} \rightarrow \alpha', a_\text{out}, 
l_\text{out}) = w_{\alpha'}\,p(\alpha', a_\text{out}^{-1}, l_\text{out} \rightarrow 
\alpha, a_\text{in}^{-1}, l_\text{in}).
\end{equation}
Taken together, this condition and the normalization conditions for the 
transition probabilities produce a system of equations from which not all 
probabilities are determined uniquely. In practice, one typically chooses an 
optimal solution according to some heuristic criteria, such as the minimization 
of “bounce” processes (those for which $\lout = \lin,~\aout = \ain^{-1}$, 
Fig.~\ref{fig:opscatter}(a))~\cite{Syljuasen2002}. Finally, it is necessary
to specify how the loop ends. If the head, carrying the action $a_h$, meets 
the tail, carrying the action $a_t$, and $a_h = a_t^{-1}$, the loop terminates. 
Otherwise, $a_t$ is replaced by the composition $a_h \circ a_t$.

So far, everything we have done is equivalent to the hand-crafted approach, 
with a slight change of nomenclature. The question for a given model is always 
how to choose the set of actions $\{a_k\}$ that the loop head can carry. In a 
sense, the most complete set of actions, $\{a_k,~ k=1,...,d-1\}$, that can be 
applied to a local basis, $\mathcal{B} = \{x_1,\dots,x_d\}$ fulfills the 
property
\begin{equation}
	\label{eq:actioncomplete}
	\{a_k(x),~ k=1,\dots,d-1\} = \mathcal{B}\backslash\{x\}\qquad 
        \forall x\in \mathcal{B},
\end{equation}
i.e., for all basis states $x \ne y$ there exists an action $a_k$ so that 
$a_k(x) = y$. In other words, the discontinuity at a loop head can change 
every state to every other state, given the right loop action. The identity 
transitions $a_k(x) = x$ can be excluded because they are in general not 
useful in changing the configuration.

Adding more actions to a complete set does not change the behavior of the 
loops, because all state discontinuities are already possible. Using a 
minimally complete set means that all hand-crafted implementations with fewer 
operators are included automatically as a solution in Eq.~\eqref{eq:detbal}. 
Such an “incomplete” set of actions would not allow all discontinuities but 
does fulfill detailed balance; a solution equivalent to the incomplete set can 
be obtained within a complete set of actions by setting the probabilities 
leading to the “forbidden” discontinuities to zero. Thus for the abstract 
directed-loop update we use a complete set of actions.

There are clearly many choices for a set of functions fulfilling the 
completeness property from Eq.~\eqref{eq:actioncomplete}. However, because all 
possible state discontinuities are already allowed, the exact implementation 
of the set does not matter. Changing from one implementation to another merely 
constitutes a relabelling of the transition indices $\alpha, \ain, \lin 
\rightarrow \alpha', \aout, \lout$. As a result, one may choose a set of 
functions on the basis of computational convenience, such as
\begin{equation}
	a^\text{mod}_k: x_n \mapsto x_{n + k \mod d},\qquad k=1,\dots, d-1,
\end{equation}
which represents the cyclic permutations of the basis states (labelled here 
starting from zero, $n = 0, \dots, d - 1$). When $d$ is a power of two, as 
in the present case, an even simpler set of functions exists in the form of 
the bitwise exclusive-or function (xor),
\begin{equation}
a^\text{xor}_k: x_n \mapsto x_{\text{xor}(n,k)},\qquad k=1,\dots, d-1.
\end{equation}
One may ask how allowing all discontinuities, as opposed to allowing only a 
selected subset, impacts computational performance. In the example of an $S = 
1$ Heisenberg model, the complete set of actions allows the head to interchange 
the states $\ket{+1}$ and $\ket{-1}$, which cannot be expressed using just the 
actions $a^+$ and $a^-$. It is well known that including $(S^\pm)^2$ loops, 
which would correspond to such discontinuities, does not increase the 
efficiency because of the $S^+_i S^-_j$ structure of the bond Hamiltonian. 
Starting from a complete set of actions, one would therefore need to go 
beyond the bounce-minimization heuristic for optimal performance and search 
for solutions of Eq.~\eqref{eq:detbal} that discourage magnetization changes 
$\Delta m \ne \pm 1$ explicitly. Using such a solution, the behavior of the 
abstract directed-loop update is again equivalent to the standard $S^\pm$ one, 
other than some performance cost associated with larger data structures.

The advantage of the abstract directed-loop update is in models and bases 
where knowledge about the optimal set of allowed discontinuities does not yet 
exist, such as the FFTL in the trimer basis. The abstract directed-loop update 
instead provides a canonical starting point, namely a complete set of actions 
with the bounce-minimization heuristic. This starting point is equivalent to 
any set of hand-crafted operators that allow all discontinuities and, in 
principle, can be optimized systematically.

In our implementation, we use the $a^\text{xor}$ actions and linear programming 
to minimize the probability of bounces in the solution of Eq.~\eqref{eq:detbal}.
While it is not always possible to avoid bounces completely, this heuristic 
alone was sufficient to achieve acceptable performance for the FFTL. More 
specifically, we were able to perform efficient simulations for systems of 
$L \times L$ trimers up to $L = 48$, at temperatures down to $T = 0.1 J_1$. For 
scans of the full $T$-$J$ plane, we restricted our simulations to systems with 
$L = 12$, this being sufficient to access the physics away from the critical 
point; close to the critical point, we performed finite-size scaling to 
capture the effect of the diverging correlation length.

\section{Thermodynamics of the FFTL}
\label{sec:fftl}

\begin{figure}
\includegraphics{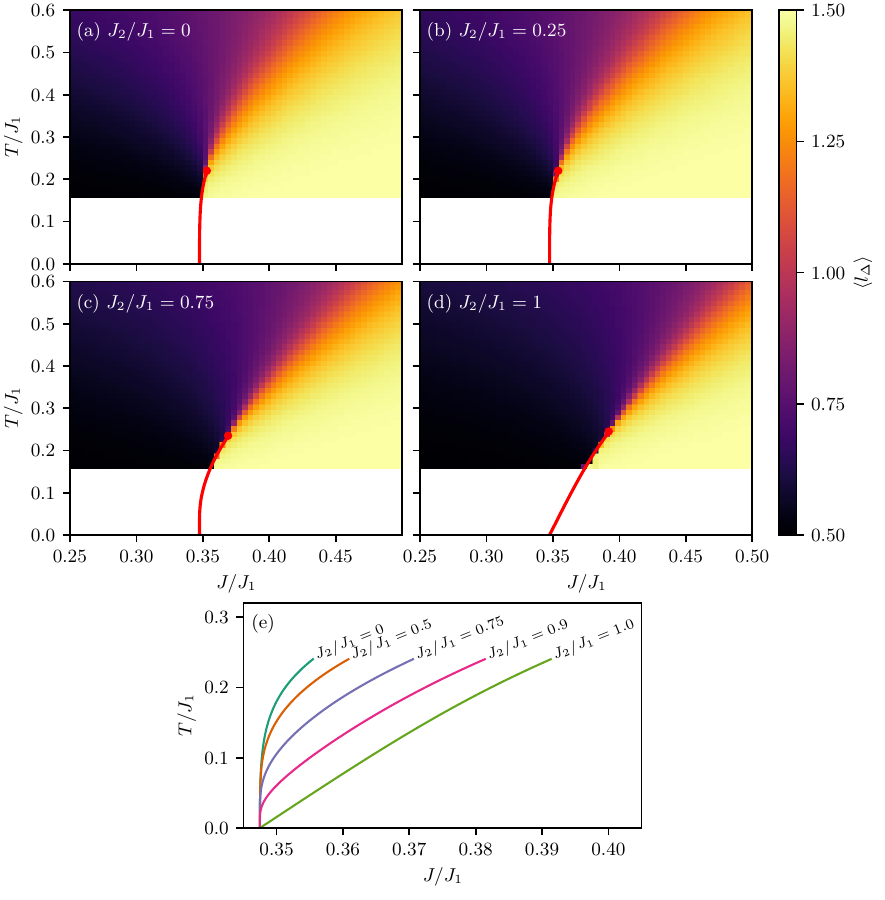}
\caption{(a)-(d) Triangle total-spin quantum number, $\langle l_\Delta \rangle$, 
computed for the $L = 12$ FFTL model at different values of the intra-trimer 
coupling ratio $J_2/J_1$, with $J_1 = J_3$. Red lines show the lines of 
first-order transitions at finite temperature, calculated from the 
approximation developed in Subsec.~\ref{sec:ffts_transition}, and 
red circles mark the critical point determined numerically in 
Subsec.~\ref{sec:ffts_critpoint}.
(e) Comparison of the shapes of the first-order lines obtained for different 
$J_2/J_1$ values from the approximation of Subsec.~\ref{sec:ffts_transition}.} 
\label{fig:ffts_phasediag}
\end{figure}

We now present the results obtained by applying the abstract directed-loop 
QMC approach to examine the thermal properties of the FFTL. Phase diagrams 
in the plane of temperature and the ratio of inter- to intra-trimer coupling 
are presented in Figs.~\ref{fig:ffts_phasediag}(a-d) for $J_1 = J_3$ with 4 
different values of $J_2/J_1$. The phase diagram is obtained from the 
expectation value of the total spin, $\mathbf{S}^2_\Delta = l_\Delta(l_\Delta+1)$, 
on each trimer and the color scale indicates the mean of its associated 
quantum number, $\braket{l_\Delta}$.

At low temperatures, $\braket{l_\Delta}$ approaches the value $1/2$ if the AFM 
intra-trimer couplings dominate, but sufficiently strong $J$ overrides this 
preference, optimizing the inter-trimer interaction terms by creating a 
ferromagnetic configuration on each trimer, causing $\braket{l_\Delta}$ to 
approach $3/2$. These effective $S = 1/2$ and $S = 3/2$ AFM ground states 
are lower in energy than any states of mixed $l_\Delta$, because the fully 
frustrated inter-trimer interactions favor neighboring trimers with the same 
$l_\Delta$. Thus they are connected directly by a level-crossing, or first-order 
quantum phase transition, occurring at one $J_2$-independent value of the 
inter-to-intra-trimer coupling ratio, $J_c/J_1$.

At low finite temperatures, this first-order transition persists, forming a line in the phase diagram. However, sufficiently high temperatures 
cause the transition to turn into a crossover, and hence each line in 
Figs.~\ref{fig:ffts_phasediag}(a-d) terminates at a critical point. This 
behavior can be associated with the effects of thermal fluctuations into 
mixed-$l_\Delta$ states and we defer a detailed determination of the critical 
points to Sec.~\ref{sec:ffts_critpoint}. 

At a superficial level these results resemble the phase diagram of the FFB, 
but there are two major distinctions. First, the first-order transition takes 
place between two phases that both have a broken symmetry at $T = 0$. This 
is in contrast to the quantum disordered dimer-singlet ground state on one 
side of the transition in the FFB. Second, in the FFTL there is another 
independent control parameter in addition to the inter-trimer coupling, and 
it is clear from Figs.~\ref{fig:ffts_phasediag}(a-d) that this $J_2/J_1$ 
parameter influences both the shape of the first-order line and the location 
of the critical point.

To investigate the physics of the FFTL phase diagram, we will first 
introduce some low-temperature approximations appropriate for gaining 
an analytical understanding of the first-order transition line 
(Subsec.~\ref{sec:ffts_transition}). Because this approach does 
not allow us to extract the position of the critical point, in 
Subsec.~\ref{sec:ffts_critpoint} we perform detailed finite-size scaling 
to identify the critical point numerically and to discuss the influence 
of $J_2$ on the critical temperature. In Subsec.~\ref{sec:ffts_specheat}, 
we examine the specific heat and discuss its evolution with the shape of 
the first-order line. In Subsec.~\ref{sec:corrlen}, we supplement this 
analysis by presenting results for the correlation length around the
critical point, which allow us to compare its line of maxima with that 
of the specific heat.

\subsection{First-order transition}
\label{sec:ffts_transition}
With the goal of finding a straightforward means of explaining the shape of 
the first-order transition line in the low-temperature region, we use the 
locally conserved quantities of the FFTL model to obtain low-temperature 
approximations to the free energy on both sides of the transition, an 
approach used for the FFB in Ref.~\cite{Stapmanns2018}. We start by 
rewriting the partition function using the local quantum numbers, 
$\{n_\Delta\} := \{(l_\Delta,l_{\Delta,13})\}$, to obtain
\begin{align}
Z &= \Tr \exp\Big[-\beta \big(\sum_\Delta H_\Delta + \underbrace{\sum_\Delta 
H_{\Delta,\Delta+\hat{x}} + H_{\Delta,\Delta+\hat{y}}}_{=: H_J}\big)\Big]\nonumber\\
&= \sum_{\{n_\Delta\}} e^{-\beta \sum_{\Delta} \varepsilon({n_\Delta})}\Tr P_{\{n_\Delta\}} 
e^{-\beta H_J} = \sum_{\{n_\Delta\}} \Tr e^{-\beta H(\{n_\Delta\})} \label{eq:splitZ}
=: \sum_{\{n_\Delta\}} e^{-\beta F(\{n_\Delta\})},
\end{align}
where $P_{\{n_\Delta\}}$ is a projection onto the subspace of the set of local 
trimer quantum numbers, $\{n_\Delta\}$, and 
\begin{equation}
H(\{n_\Delta\}) = \sum_\Delta \varepsilon(n_\Delta) + P_{\{n_\Delta\}} H_J P_{\{n_\Delta\}}.
\end{equation}
Thus, $F(\{n_\Delta\})$ is the free energy of a system with a frozen 
configuration of trimer quantum numbers belonging to the partition function
\begin{equation}
    Z(\{n_\Delta\})=\Tr e^{-\beta H(\{n_\Delta\})}.
\end{equation}
In this way, the FFTL model can be regarded as a mixture of all possible ways 
to combine $S = 1/2$ and $S = 3/2$ trimer spins into a square-lattice AFM.
Because $F(\{n_\Delta\})$ is an extensive quantity, the distribution in 
Eq.~\eqref{eq:splitZ} will generally become sharp in the thermodynamic limit 
and at low temperature, and the system will choose the configuration 
$\{n_\Delta\}$ that minimizes $F(\{n_\Delta\})$.

Our first approximation is that $F(\{n_\Delta\})$ is minimized by a uniform 
configuration, $n_\Delta = n$. At low temperatures, this is expected because 
a system containing multiple domains of different $n_\Delta$ can decrease its 
energy by growing the single domain with the lowest energy per site. At high 
temperatures, however, “domain walls” become favorable due to their entropy, 
and lead to the appearance of the critical point. By neglecting domain walls, 
our approximation loses the ability to predict the end of the first-order 
line. Further, it is conceivable that domain walls between $l_\Delta = 1/2$ and 
$l_\Delta = 3/2$ could become energetically favorable even at $T = 0$, due to 
quantum fluctuation effects, although we will exclude this possibility 
\textit{a posteriori} for the bare FFTL by comparing our approximate 
results to the QMC data.

For simplicity, we first consider $T = 0$, where $F(\{n_\Delta\})$ reduces to 
the ground-state energy, $E_0(\{n_\Delta\})$, of $H(\{n_\Delta\})$. As explained 
above, we assume
\begin{equation}
    \min_{\{n_\Delta\}} E_0(\{n_\Delta\}) = E_0(n)
\end{equation}
and restrict our considerations to three possible ground-state configurations 
with uniform $n_\Delta$, having total energies per trimer ($N_t$ denotes the 
number of trimers)
\begin{align}
E_\mathrm{Q}/N_t &= \varepsilon_\mathrm{Q} + J \varepsilon_\text{AFM}^{S=3/2},\\
E_{\mathrm{D}1}/N_t &= \varepsilon_{\mathrm{D}1} + J \varepsilon_\text{AFM}^{S=1/2},\\
E_{\mathrm{D}0}/N_t &= \varepsilon_{\mathrm{D}0} + J \varepsilon_\text{AFM}^{S=1/2},
\end{align}
where $\varepsilon_\text{AFM}^{S=1/2}$ ($\varepsilon_\text{AFM}^{S=3/2}$) is the 
ground-state energy per site of the spin-1/2 (spin-3/2) square-lattice 
Heisenberg AFM with unit couplings. We denote by $\mathrm{D0}$ the doublet 
with lower energy, $\varepsilon_{\mathrm{D}0}\le \varepsilon_{\mathrm{D}1}$, to 
obtain for the ground state
\begin{align}
    n = \begin{cases} \mathrm{D}0 & \text{if } \varepsilon_{\mathrm{D}0}
    + J \varepsilon^{S=1/2}_\text{AFM} \le \varepsilon_{\mathrm{Q}}
    + J \varepsilon^{S=3/2}_\text{AFM},\\
    \mathrm{Q} & \text{otherwise.}
    \end{cases}
\end{align}
The critical inter-trimer coupling where the two levels cross is 
\begin{equation}
\label{eq:jczerot}
    J_c = \frac{\varepsilon_{\mathrm{Q}}-\varepsilon_{\mathrm{D}0}}
    {\varepsilon^{S=1/2}_\text{AFM} - \varepsilon^{S=3/2}_\text{AFM}},
\end{equation}
and because $J_c > 0$ a first-order quantum phase transition takes place 
between a spin-1/2 and a spin-3/2 AFM phase. The value of $J_c$ does not 
depend on $J_2$ because $l_{\Delta,13} = 1$ in both the Q and D0 levels, 
and the contribution from the $J_2$ dimer cancels in the numerator of 
Eq.~\eqref{eq:jczerot}. 

The reasoning to this point is analogous to the FFB case, with the exception 
of the additional quantum number, $l_{\Delta,13}$, that labels the doublets 
$\mathrm{D}0$ and $\mathrm{D}1$. Although the $\mathrm{D}1$ level has no 
direct influence at $T = 0$, this changes at finite temperatures. When $T > 0$, 
the first-order transition is expected to continue along a line, $J_c(T)$, up 
to the critical point at $(T_c, J_c(T_c))$. For $T \ll T_c$, we extend the 
approximation to finite temperatures by replacing energies with free energies. 
For non-degenerate $\mathrm{D}0$ and $\mathrm{D}1$, the system will adopt the 
state with the lower free energy of
\begin{align}
    F_{\mathrm{Q}}/N_t &= \varepsilon_\mathrm{Q} + J \varepsilon_\text{AFM}^{S=3/2}
    (T/J) - T s_\text{AFM}^{S=3/2}(T/J),\\
    F_{\mathrm{D}0}/N_t &= \varepsilon_{\mathrm{D}0} + J \varepsilon_\text{AFM}^{S=1/2}
    (T/J) - T s_\text{AFM}^{S=1/2}(T/J).
\end{align}
However, if $\epsilon_{\mathrm{D}1} - \epsilon_{\mathrm{D}0} \lesssim T$, it is 
not meaningful to consider the configuration $n_\Delta = \mathrm{D}0$ as a 
macroscopic state, because thermal fluctuations allow admixtures of 
$\mathrm{D}0$ and $\mathrm{D}1$ to acquire a finite weight. Instead we 
subsume all configurations with $n_\Delta\in\{\mathrm{D}0,\mathrm{D}1\}$ into 
a single macrostate,
\begin{align}
    e^{-\beta F_\mathrm{D}} &= \sum_{\substack{\{n_\Delta\}\\n_\Delta= \mathrm{D}0,\mathrm{D}1}} 
    e^{-\beta [\varepsilon_n + F(\{n_\Delta\})]} = \left(1 + e^{-\beta (\varepsilon_{\mathrm{D}1}
     - \varepsilon_{\mathrm{D}0})}\right)^{N_t} e^{-\beta F_{\mathrm{D}0}},
\end{align}
with free energy
\begin{align}
F_\mathrm{D} &= F_{\mathrm{D}0} - T S_\mathrm{D},\\
S_\mathrm{D} &= N_t \log\left(1+e^{-\beta(\varepsilon_{\mathrm{D}1} - \varepsilon_{\mathrm{D}0})} 
\right).
\end{align}
As expected, the additional entropy contribution due to the doublets vanishes 
when $\varepsilon_{\mathrm{D}1} - \varepsilon_{\mathrm{D}0} \gg T$ and assumes the 
maximal value of $N_t \log 2$ in the degenerate case $\varepsilon_{\mathrm{D}1} = 
\varepsilon_{\mathrm{D}0}$ ($J_1 = J_2 = J_3$).

These expressions allow us to estimate the first-order line, $J_c(T)$, for 
$T \ll T_c$ in a manner similar to Ref.~\cite{Stapmanns2018}. Because 
$F_\mathrm{Q}$ and $F_{\mathrm{D}0}$ describe square-lattice Heisenberg AFMs, we 
employ the standard low-temperature approximation for the specific heat,
$C(T,J) = \partial E(T,J)/\partial T/N_t$, to capture the behavior of the 
free energy at low temperatures,
\begin{align}
F(T,J) &= E(0,J) - T S(0,J) + \int_0^T dT'~\left(1-\frac{T}{T'}\right) N_t 
C(T',J).
\end{align}
For the spin-$S$ AFM, the spin-wave excitations (magnons) give a power-law 
contribution to the specific heat~\cite{Hasenfratz1993, PhysRevB.47.7961, 
PhysRevB.50.6877, Johnston2011}, 
\begin{align}
\label{eq:specheat}
    C(T,J) &= \frac{\partial J \varepsilon_\text{AFM}(T/J)}{\partial T}
    = a(S) \left(\frac TJ\right)^2 + b(S) \left(\frac TJ\right)^4 + \dots,
\end{align}
where we have also included the next-to-leading-order correction. The free 
energy then becomes
\begin{equation}
\label{eq:free_energy_integral}
F(T) = E(0)- T S(0) - \frac{a(S)}{6J^2} T^3 - \frac{b(S)}{20 J^4} T^5 + \dots,
\end{equation}
or for the FFTL model 
\begin{align}
F_\mathrm{Q}(T,J)/N_t &= \epsilon_\mathrm{Q} + J\varepsilon_\text{AFM}^{S=3/2}
 - \frac{a(3/2)}{6J^2} T^3 - \frac{b(3/2)}{20 J^4} T^5,\\
F_{\mathrm{D}0}(T,J)/N_t &= \epsilon_{\mathrm{D}0} + J\varepsilon_\text{AFM}^{S=1/2}
 - \frac{a(1/2)}{6J^2} T^3 - \frac{b(1/2)}{20 J^4} T^5,\\
F_{\mathrm{D}}(T,J)/N_t &= F_{\mathrm{D}0}(T)/N_t - T S_\mathrm{D}/N_t.
\label{eq:fd_sd}
\end{align}
Values for $a(1/2) = 0.8252$, $a(3/2) = 0.1146$, $b(1/2) = 7.2$, and 
$b(3/2) = 0.067$ are provided by Ref.~\cite{Johnston2011} and the value  
$\varepsilon_\text{AFM}^{S=1/2}=-0.669437(5)$ by Ref.~\cite{Sandvik1997}. We 
did not find a value for $\varepsilon_\text{AFM}^{S=3/2}$ in the literature, 
and so we extracted the result $\varepsilon_\text{AFM}^{S=3/2} = -4.98603(3)$ 
from our own SSE QMC simulations, as detailed in App.~\ref{sec:app_gs_energy}.

Solving the equation
\begin{align}
    \label{eq:foline_d}
    F_\mathrm{Q}(T,J_c) &= F_\mathrm{D}(T,J_c)
\end{align}
for different, fixed values of $J_2/J_1$ then yields the transition lines, 
$J_c(T)$, shown in Fig.~\ref{fig:ffts_phasediag}(e). Direct comparison with 
the data shown for a finite-sized system in Figs.~\ref{fig:ffts_phasediag}(a-d) 
confirms that the approximate treatment holds remarkably well in describing the 
first-order transition line. In fact, within the finite grid of our $(J,T)$ 
coverage, we do not detect any deviations from it, even up to $T_c$. However, 
as discussed earlier, the existence of the critical point is not visible by 
this approach.

The approximate analysis provides valuable insight into the shapes of the 
transition lines in Fig.~\ref{fig:ffts_phasediag}(e). In the limit of low 
temperatures and when $J_2 \ne J_1$, $S_D$ is suppressed exponentially and 
$J_c(T)$ is dominated by the $T^3$ magnon contribution, leading to the 
asymptotic behavior
\begin{equation}
J_c(T) \sim J_c + \frac{\varepsilon_\text{AFM}^{S=3/2}
 - \varepsilon_\text{AFM}^{S=1/2} }{ 6(\epsilon_\mathrm{Q}
 - \epsilon_\mathrm{D0})^2}\left( a(3/2) - a(1/2)\right)\,T^3,
\end{equation}
which describes an asymptotically vertical transition line. However, in the 
special case $J_1 = J_2 = J_3$, $S_D = N_t \log 2$ never becomes irrelevant 
due to the extensive ground-state degeneracy in the doublet phase. Because 
$- T S_D$ enters linearly, the result is an asymptotically sloped first-order 
line,
\begin{equation}
J_c(T) \sim J_c + \frac{\log 2}{\varepsilon_\text{AFM}^{S=1/2}
 - \varepsilon_\text{AFM}^{S=3/2}}\,T, 
\end{equation}
accounting for the clear qualitative difference in lineshapes 
in Fig.~\ref{fig:ffts_phasediag}(e). A closer inspection of 
Eq.~\eqref{eq:free_energy_integral} reveals that obtaining a 
$T$-linear free-energy contribution, and thus a first-order line 
with finite slope at $T = 0$, is possible only in the presence 
such a macroscopic jump in the ground-state entropy. 

For intermediate $J_2/J_1$ and at finite temperatures, the doublet entropy 
term, $-T S_D$, competes with the magnon terms, leading to a smooth crossover 
between the regime of well separated doublets and the case where they are 
degenerate (Fig.~\ref{fig:ffts_phasediag}(e)). Even close to $T_c$, the slope 
of the first-order line is influenced by this competition, allowing us to 
alter the slope at $T_c$ to a certain extent by changing $J_2/J_1$. We will 
explore the effects of this parameter in Subsecs.~\ref{sec:ffts_specheat}
and~\ref{sec:corrlen}. However, first we focus on locating the critical point, 
which is not possible within the approximate analytical approach, and for 
this we return to QMC simulations and systematic finite-size scaling. 

\subsection{Thermal critical point}
\label{sec:ffts_critpoint}

\begin{figure}
\includegraphics{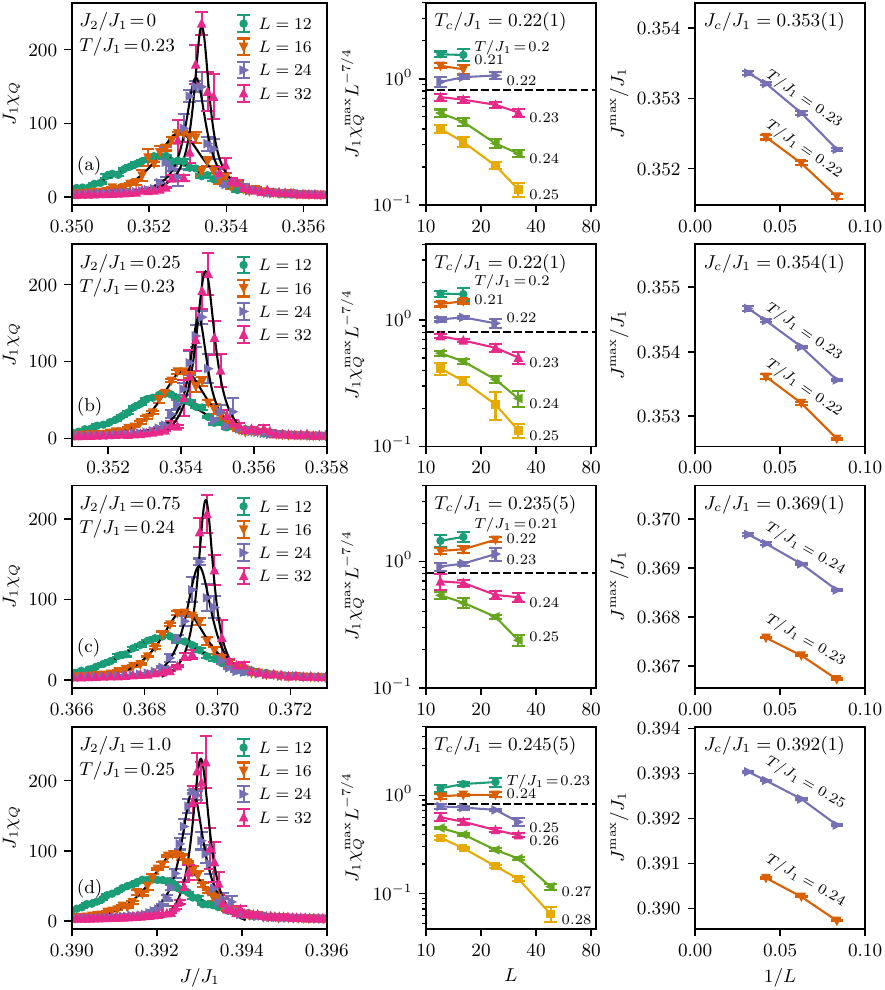}
\caption{Determination of the critical point of the FFTL model for different 
intra-trimer couplings, $J_2/J_1$, by finite-size scaling of the maximum in 
the quartet susceptibility, $\chi_Q(T,J)$, computed along several fixed-$T$ 
cuts. (left column) $\chi_Q(T \approx T_c, J)$ for different $L$. The maxima 
are fitted using a Lorentzian (black). (center column) Scaling of the maximum 
value, $\chi^\text{max}_Q$, normalized by the power-law form, $L^{\gamma/\nu} = 
L^{7/4}$, expected for the 2D Ising universality class. (right column) Scaling 
of the position, $J^\text{max}$, of this maximum.}
\label{fig:ffts_critpoint}
\end{figure}

As we saw in Fig.~\ref{fig:ffts_phasediag}, the first-order transition line, 
$J_c(T)$, continues only up to a critical temperature, $T_c$. Because the 
fluctuations that proliferate at this critical point are those of the scalar 
$l_\Delta$ degree of freedom, one expects by analogy with the phase diagram of 
water, and of the FFB~\cite{Stapmanns2018} and SSL~\cite{Jimenez2020}, a 2D 
Ising-type criticality in the FFTL. We now extract $(T_c,J_c(T_c))$ for the 
different values of $J_2/J_1$ shown in Fig.~\ref{fig:ffts_phasediag}.

Because both the critical temperature, $T_c$, and the critical coupling, 
$J_c(T_c)$, are unknown, locating the critical point numerically requires 
the adjustment of two parameters. To accomplish this, we calculated the 
“quartet susceptibility,”
\begin{align}
\chi_{Q} = \frac{\partial^2 f}{\partial \varepsilon_{Q}^2} 
&= \frac{\beta}{N_t} \left(\Braket{N_Q^2}- \Braket{N_Q}^2\right),\\
&= \frac{\beta}{N_t} \left(\Braket{\left(\sum_{\Delta}l_\Delta\right)^2}- 
\Braket{\sum_{\Delta} l_\Delta}^2\right),
\end{align}
along lines of fixed $T$. $\chi_{Q}$ reflects the fluctuations in the 
number of $l_\Delta = 3/2$ sites, $N_Q$, or equivalently fluctuations in the 
sum over all $l_\Delta$, and diverges along the first-order line, $J_c(T)$ 
with $T \le T_c$, whereas for $T > T_c$ the divergence turns into a finite 
peak. In a finite system, any divergence at a phase transition is regularized 
by the system size and also appears as a finite peak. We therefore search for 
the critical point along the line of susceptibility maxima, $(T, J^\text{max} 
(T))$, where the peak height of the susceptibility will follow a (divergent) 
power-law form as a function of the system size,
\begin{equation}
    \chi^\text{max}_{Q}(T_c) \sim L^{\gamma/\nu},
\end{equation}
with the critical exponents $\gamma$ and $\nu$. If we assume a 2D Ising 
critical point, we can find its position by ensuring the best match between 
the finite-size scaling of the maximum and the exactly known exponents 
$\gamma = 7/4$ and $\nu = 1$~\cite{Fisher1967}.

In Fig.~\ref{fig:ffts_critpoint}, this program is carried out for different 
values of $J_2/J_1$. In practice, the simulations become less efficient close 
to the first-order transition, making the extraction of the maximum, 
$\chi_Q^\text{max}$, intractable for larger system sizes and $T \lesssim T_c$. 
Nevertheless, by using the known 2D Ising exponents, we are able to extract 
accurate estimates. Their error is dominated by the uncertainty in $T_c$, 
which is mostly a consequence of the maximum system sizes we can simulate. 
We find that $J_c(T_c)$ rises significantly, whereas $T_c$ itself is enhanced 
only weakly, as $J_2/J_1$ approaches $1$. Marking the positions of these 
critical points in the phase diagrams of Figs.~\ref{fig:ffts_phasediag}(a-d), 
we find again remarkable agreement between the QMC data and the analytical 
estimate for the first-order line: the fact that the critical points fall 
on top of the estimated lines shows that the approximations of 
Subsec.~\ref{sec:ffts_transition} perform well even up to criticality. 
The rise of $J_c(T_c)$ with $J_2/J_1$ reflects the increasing slope of 
the first-order lines with $J_2/J_1$, whose thermodynamic consequences 
we consider next. 

\subsection{Specific heat}
\label{sec:ffts_specheat}

The critical point of the FFTL, and of similar models such as the FFB and 
SSL~\cite{Stapmanns2018,Jimenez2020}, can be viewed up to the differing 
dimensionality ($d = 2$ vs. $d = 3$) as a relative to the critical point 
that terminates the liquid-gas transition line of water. Indeed, these spin 
models share the critical exponents of the $d = 2$ Ising universality class, 
in the same way that the critical point of water shows $d = 3$ Ising 
universality. Further, the critical point in these quantum spin models comes 
about without the presence of an explicit $Z_2$ symmetry, making them even 
closer relatives of water than is the $d = 3$ Ising model itself. 

At a second glance, however, this correspondence is not trivial. An immediate 
example arises in a comparison of the specific heat that was performed 
recently~\cite{Jimenez2020}. In the quantum spin systems, the specific-heat 
data contain two lines of maxima that both approach the critical point sideways 
relative to the first-order line.
While one line of maxima is well known in the specific heat of water, appearing as
an extension of the first-order line beyond the critical point and into the supercritical
regime, there is no equivalent to the second line of maxima seen in the spin systems.
Another distinction is the slope of the first-order 
line at the critical point with respect to the temperature axis, the FFB and 
SSL models featuring rather vertical lines quite opposed to the strongly 
slanted one in water (explicit visualizations may be found in 
Ref.~\cite{Jimenez2020}). These differences raise the general question of 
the extent to which the two lines of maxima can be a universal feature of 
the critical point, rather than a model-specific property, and we would like 
in particular to understand their connection with the slope of the first-order 
line.

\begin{figure*}
\includegraphics{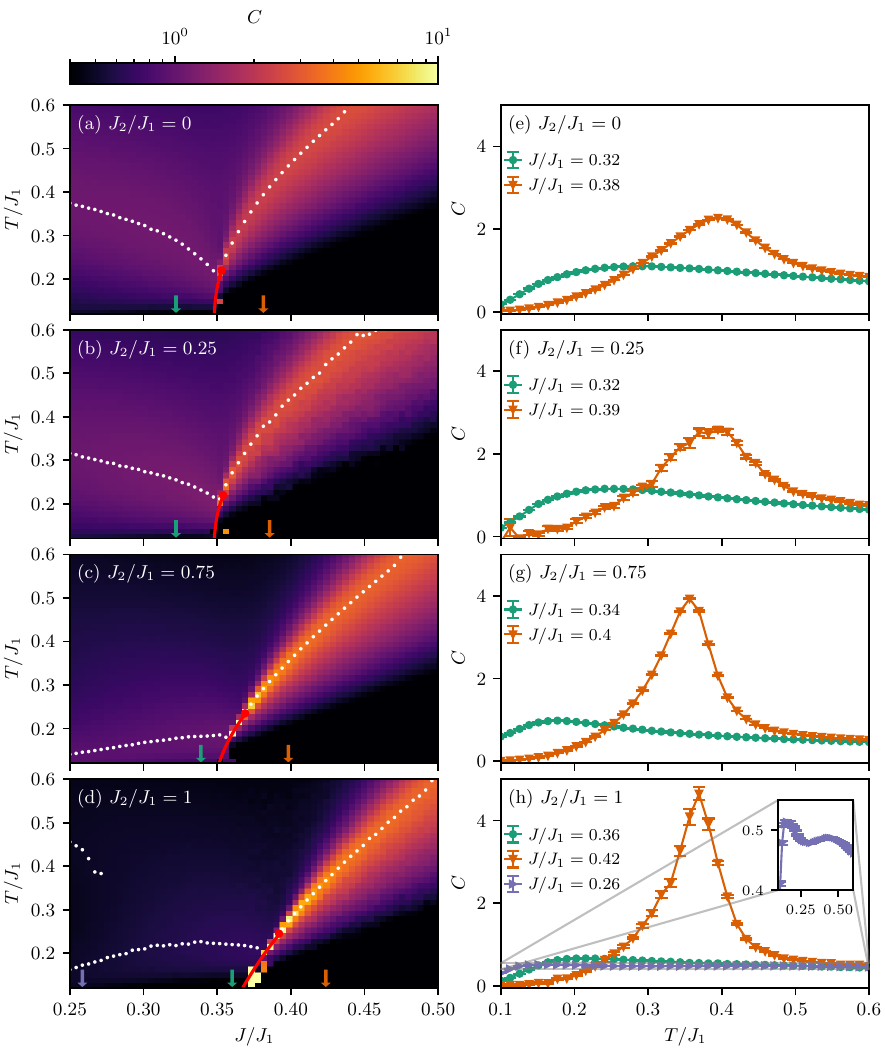}
\caption{Specific heat, $C(J,T)$, of the $L = 12$ FFTL model for different 
values of $J_2/J_1$. (a)--(d) $C$ as a function of $J$ and $T$. For each fixed 
$J$, the temperatures, $T^\text{max}$, of the maxima in $C(T)$ are shown as 
white dots. Red lines and points mark again the approximate first-order line 
and the numerically determined critical point. Arrows indicate the positions 
of the fixed-$J$ cuts shown in panels (e)--(h). In panel (h), the inset shows 
a magnification of the cut at $J/J_1 = 0.26$, highlighting the two-maxima 
structure.}
\label{fig:ffts_specheat_scan}
\end{figure*}

\begin{figure}
  \centering
  \includegraphics{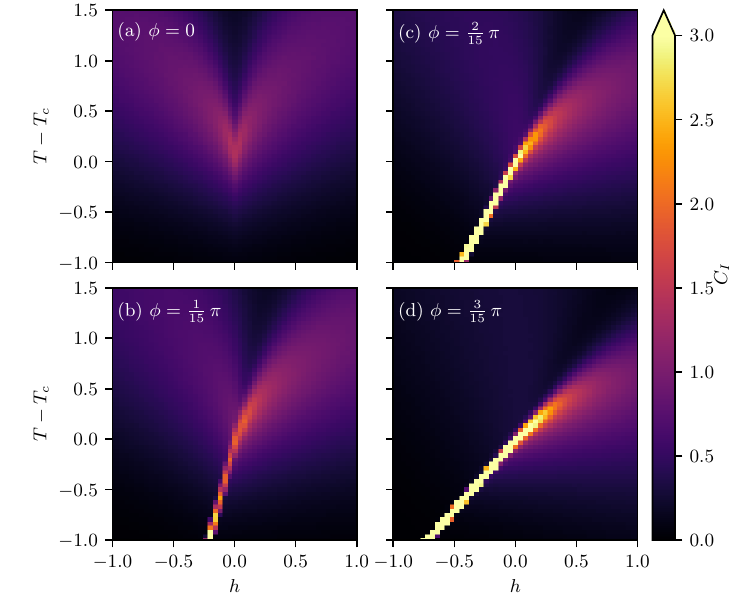}
  \caption{Specific heat, $C_I(T,h)$, of the bare $L = 12$ Ising model 
  in rotated coordinates, shown for different values of the mixing
  angle, $\phi$, between the bare coordinates $\tilde{T}$ and $\tilde{h}$.}
  \label{fig:ising}
\end{figure}

The FFTL, in which $J_2/J_1$ produces quite explicitly the differently slanted 
first-order transition lines shown in Fig.~\ref{fig:ffts_phasediag}(e), thus 
provides a valuable model for elucidating this connection. In 
Fig.~\ref{fig:ffts_specheat_scan} we show the specific heat, $C = \partial 
\langle H \rangle /\partial T$, in the $T$-$J$ plane for 4 different values 
of $J_2$. In the vicinity of the critical point, we indeed identify two lines 
of maxima, whose shapes depend on the strength of $J_2$. Although neither the 
lines nor the actual $C(T)$ curves are ever symmetrical, for $J_2 = 0$ 
(Figs.~\ref{fig:ffts_specheat_scan}(a,e)) the lines of maxima do appear at 
comparable temperatures. This is similar to the specific-heat results for the 
FFB and SSL models~\cite{Jimenez2020}, in which we stress that there is no 
exact symmetry, whereas this symmetry is an exact property of the Ising 
model (enforced by the $Z_2$ symmetry of the Hamiltonian and displayed in 
Fig.~\ref{fig:ising}(a)). 

As $J_2$ is increased and the first-order line slants further to the right, 
both the temperatures (Figs.~\ref{fig:ffts_specheat_scan}(b-d)) and peak 
heights (Figs.~\ref{fig:ffts_specheat_scan}(f-h)) of the specific heat are 
suppressed on the left side of the phase diagram ($S = 1/2$ AFM). At $J_2/J_1
 = 1$ (Fig.~\ref{fig:ffts_specheat_scan}(d,h)), the maxima to the left of 
$J_c$ are barely discernible, least of all when compared to the robust peak 
in their right-side counterparts, and the situation resembles the specific 
heat of water, where only one line of maxima is visible. In this way, the FFTL 
provides a continuous progression from the “Ising-like” to the “water-like” 
form of the specific heat within a single model, making it a useful example 
to understand their differences. At the microscopic level, it is the 
doublet-degeneracy of the FFTL model that allows us to control this 
progression, and to proceed all the way to the asymmetric, “water-like” form 
in a manner that was not possible in the FFB and SSL models (which are not 
far from “Ising-like” in this respect). 

Focusing in detail on the positions of the maxima, we observe that only the 
right-hand line of maxima approaches the critical point directly. The left-hand 
line of maxima instead meets the first-order line below the critical temperature, 
$T_c$. This behavior is most obvious around $J_2/J_1 = 1$ and becomes harder to 
distinguish for smaller $J_2$. However, its presence does point to the fact 
that the left-hand maxima are not a universal feature of the critical point, 
and thus it is not surprising that there are models where only one line of 
maxima appears.

Returning to the analysis of Subsec.~\ref{sec:ffts_transition}, the specific 
heat of the FFTL for $J < J_c$ can be understood as the sum of an $S = 1/2$ 
AFM magnon contribution and the specific heat of decoupled trimers ($J = 0$). 
The magnon contribution consists of a peak at the temperature scale of $J$, 
which is independent of the value of $J_2$. Conversely, the trimer contribution 
is independent of $J$ but depends strongly on $J_2$. When the degeneracy 
between $\epsilon_{\mathrm{D}0}$ and $\epsilon_{\mathrm{D}1}$ is broken, the 
$\mathrm{D}1$ level moves upwards into the gap between $\epsilon_{\mathrm{D}0}$
and $\epsilon_\mathrm{Q}$, leading for small $J_2/J_1$ to a “doublet peak” that 
obscures the magnon peak. As $J_2/J_1 \rightarrow 1$, the gap between the 
doublet levels closes and the associated specific-heat peak moves to lower 
temperatures, revealing more of the magnon peak. In the degenerate case 
($J_2/J_1 = 1$), the doublet entropy is not released at all, leading to a 
very weak specific-heat signal. In this situation, the magnon and trimer 
contributions are sufficiently well separated at weak inter-trimer couplings 
($J/J_1 \lesssim 0.27$) that they produce two clearly distinguishable, if 
very weak, local maxima (inset, Fig.~\ref{fig:ffts_specheat_scan}(h)).

From another perspective, one may ask how much of the specific heat can be 
understood by considering the free energy close to the critical point, where 
its singular part will become similar to the part associated with the 2D 
Ising universality class. This similarity, however, holds only up to a 
coordinate transformation in the scaling fields, i.e.~the quantities known 
as the temperature and magnetic field in the bare Ising model need not 
correspond to the physical temperature and couplings of the FFTL.

This type of coordinate transformation not only controls the angle and 
location of the first-order line in the $(J,T)$ phase diagram, but also 
has a direct effect on the thermodynamic quantities by transforming and 
mixing their free-energy derivatives. To elucidate this point, we performed 
such a transformation explicitly for the specific heat of the bare Ising model 
on the square lattice,
\begin{equation}
H = -\sum_{\braket{i,j}} \sigma^z_i \sigma^z_j - \tilde{h} \sum_i \sigma^z_i,
\end{equation}
at “bare” temperature $\tilde{T}$ and magnetic field $\tilde{h}$. To compare 
with the FFTL, which has a variable slanting of the first-order line, we 
introduce a coordinate transformation that is essentially a rotation around 
the critical point, 
\newcommand{\vect}[1]{\begin{pmatrix}#1\end{pmatrix}}
\begin{equation}
	\vect{T-T_c\\h} = \vect{\cos(\phi) & -\sin(\phi)\\\sin(\phi) & 
        \phantom{-}\cos(\phi)} \vect{\tilde{T}-T_c\\\tilde{h}};
\end{equation}
(a related transformation was considered in the context of liquid-liquid 
transitions in Ref.~\cite{Luo2014}). Under this transformation, the specific 
heat mixes with the magnetocaloric correlations and the magnetic susceptibility 
to give
\begin{equation}
\label{eq:fieldmixing}
- C_I/T = \frac{\partial^2 f}{\partial T^2} = \cos^2(\phi)\,\frac{\partial^2 f}
{\partial \tilde{T}^2} + \cos(\phi) \sin(\phi)\,\frac{\partial^2 f}{\partial 
\tilde{T}\partial\tilde{h}} + \sin^2(\phi) \, \frac{\partial^2 f}{\partial
\tilde{h}^2}.
\end{equation}

We evaluate these three free-energy derivatives for the Ising model by 
computing the correlation functions $\braket{H^2} - \braket{H}^2$, $\braket{H 
M} - \braket{H}\!\braket{M}$, and $\braket{M^2} - \braket{M}^2$, where $M = 
\sum_i \sigma_i^z$ is the Ising magnetization. For this purpose we employed 
classical Monte Carlo simulations with Metropolis~\cite{Metropolis1953} and 
Wolff~\cite{Wolff1989} updates using the “ghost spin” method~\cite{Dobias2018} 
to simulate efficiently in a magnetic field. Figure~\ref{fig:ising} shows the 
“rotated” specific-heat results we obtained for different values of the 
coordinate mixing angle, $\phi$. Looking again at the maxima, we observe 
that singular contributions appear along the first-order line as soon as 
$\phi > 0$, and that the right-hand maxima above it are enhanced. The more 
slanted the first-order line becomes, the more strongly is the right branch 
enhanced, until even at comparatively small $\phi$ values it dominates the 
signal completely. 

This picture is in good agreement with the specific heat of the FFTL close 
to the critical point in Fig.~\ref{fig:ffts_specheat_scan}, where all cases 
feature slanted first-order lines whose angle increases with their proximity 
to the degenerate case. Similar behavior is also present in the specific-heat 
data for the FFB and SSL models~\cite{Jimenez2020}. At this point, we stress 
that the construction of the “rotated” free energy has a meaning only in the 
vicinity of the critical point: it cannot be written in terms of a partition 
function of a microscopic model and away from criticality it is the 
microscopic degrees of freedom that are the most important. Specifically for 
our discussion of the full first-order lines, their low-temperature shapes are 
those given by the free-energy arguments of Subsec.~\ref{sec:ffts_transition} 
and shown in Fig.~\ref{fig:ffts_phasediag}(e).

\begin{figure}
	\centering
	\includegraphics{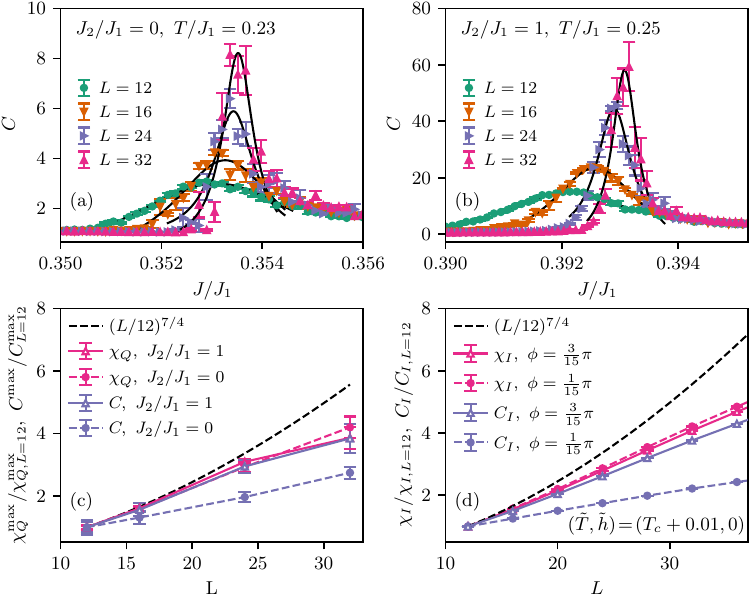}
	\caption{Finite-size scaling of the specific heat, $C(J,T)$, close to the
	critical point for (a) a non-degenerate FFTL and (b) the degenerate FFTL. (c)
	Scaling of the specific-heat maxima, $C^\text{max}$, shown together with the
	quartet susceptibility maxima, $\chi_Q^\text{max}$, normalized for comparison
	to their $L = 12$ values. The dashed black line shows a power law with the
	critical exponent $\gamma/\nu = 7/4$. The temperatures are those of panels
	(a) and (b). (d) Scaling of the specific heat, $C_I$, and the susceptibility,
	$\chi_I$, in the rotated Ising model at different mixing angles, $\phi$,
	normalized to their values with $L = 12$. The data are taken with a small
	offset from the critical point to mimic the situation in panel (c). }
	\label{fig:ffts_specheat}
\end{figure}

The physics underlying these results is the fact that the most divergent
quantity in the critical scaling of the Ising model is the susceptibility,
given by the third term of Eq.~\eqref{eq:fieldmixing}, whereas the Ising
specific heat (first term) has only a weak critical scaling ($\log L$ in
2D). The critical properties of any system described by a rotated Ising
model are a mixture of the pure Ising terms, and can therefore be expected
to change their form. This behavior is seen most clearly in the critical
specific heat of the FFTL, which in contrast to the 2D Ising form grows
strongly with system size, as we show in Figs.~\ref{fig:ffts_specheat}(a-b).
While it is difficult to confirm the true asymptotic critical scaling in our
FFTL-model data due to the limited accuracy with which we can identify the
exact critical point, some useful observations are nevertheless possible.
In the degenerate case ($J_2/J_1 = 1$), the scaling of the specific heat
coincides with that of the quartet susceptibility up to a normalization
factor (Fig.~\ref{fig:ffts_specheat}(c)). For $J_2/J_1 = 0$, the specific
heat does grow more slowly than the quartet susceptiblity, in agreement
with the weaker scaling-field mixing of the non-degenerate case, although
the susceptibility contribution is still expected to dominate at large
system sizes.

A purely Ising-type $\log L$ scaling of the specific heat has been observed
at the thermal critical point in the FFB \cite{Stapmanns2018}, indicating
that the mixing of scaling fields is weaker in that model than in the maximally
non-degenerate FFTL. For the degenerate FFTL, the match in the scaling behavior
of the specific heat and the quartet susceptibility close to critical point
can be interpreted thermodynamically as an effect of scaling-field mixing
and microscopically by the fact that the energy fluctuations (reflected
in $C$) are dominated by quartet fluctuations ($\chi_Q$) in the absence
of a gap between the doublets (Fig.~\ref{fig:ffts_specheat_scan}(h)). This
interpretation is further confirmed by a direct comparison of the model
results (Fig.~\ref{fig:ffts_specheat}(c)) with the scaling of the specific
heat in the rotated Ising model (Fig.~\ref{fig:ffts_specheat}(d)) for
different mixing angles. The deviation from the critical $L^{\gamma/\nu}$ scaling
in both cases is due to the limited resolution in $(J,T)$ with which we are
able to identify the critical point in the FFTL, a situation we introduce in
Fig.~\ref{fig:ffts_specheat}(d) by a deliberate small detuning from the Ising
critical temperature, $T_c$.

Finally, in principle one may question the effect of this 
scaling-field mixing on our estimation of the critical point in 
Subsec.~\ref{sec:ffts_critpoint}, where we compared the growth of the 
quartet susceptibility directly to the critical scaling of the Ising 
susceptibility.
However, these two quantities no longer have the same direct relationship when
 the scaling fields become mixed: similar to 
Eq.~\eqref{eq:fieldmixing}, the scaling of the singular part of the 
quartet susceptiblity is given by a combination of Ising observables. 
However, once again the most divergent contribution is that of the 
Ising susceptibility, which therefore dominates in the thermodynamic 
limit, while the other mixed-in contributions behave as subleading
corrections to scaling and enter our estimate of the critical point 
in the same way as additional finite-size effects. The contributions from 
weaker divergences can be significant in some quantities for the 
system sizes of our study, most notably for $C(T)$ at smaller values 
of $J_2/J_1$. However, the results of Fig.~\ref{fig:ffts_specheat}(c) 
make clear that the quartet susceptibility, $\chi_Q$, of the FFTL is 
affected only weakly by the mixing of scaling fields for all values 
of $J_2/J_1$, making it eminently suitable for a robust extraction of 
$T_c$ and $J_c(T_c)$. In fact the minor deviation from $L^{\gamma/\nu}$ 
behavior of $\chi_Q$ is most likely the consequence of a small 
discrepancy from the exact critical point, which would allow our 
estimate to be further improved. 

\begin{figure}
\centering
\includegraphics{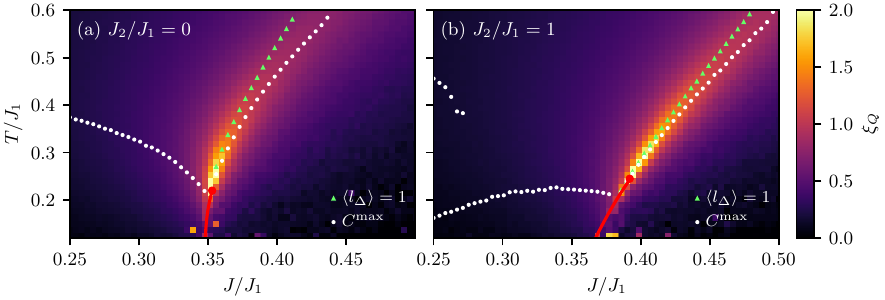}
\caption{Quartet correlation length, $\xi_Q$, shown as a function of $T$ 
and $J$ for the $L = 12$ FFTL model at (a) a representative non-degenerate 
intra-trimer coupling, $J_2/J_1 = 0$, and (b) the degenerate point, $J_2/J_1
 = 1$. Superimposed are the lines of constant-$J$ maxima in the specific heat, 
$T^\text{max}$, and the “critical isochore” line where $\braket{l_\Delta} = 1$.
In both panels, the critical point and the approximate first-order line are 
shown in red.}
\label{fig:correlen}
\end{figure}

\subsection{Correlation length and characteristic lines}
\label{sec:corrlen}

Additional insight into the physics above $T_c$ can be obtained by 
considering the correlation length associated with the critical point. 
This we estimate~\cite{Sandvik2010} from the quartet structure factor,
\begin{equation}
S_Q(\mathbf{q}) = \frac{1}{N_t} \sum_{\Delta,\Delta'} e^{-i \mathbf{q} \cdot 
(\mathbf{R}_\Delta-\mathbf{R}_{\Delta'})} \left(\braket{l_\Delta l_{\Delta'}}
 - \braket{l_\Delta}\!\braket{l_{\Delta'}}\right),
\end{equation}
as
\begin{equation}
\def\dq{\delta\mathbf{q}}
\xi_Q = \frac{1}{|\dq|} \sqrt{\frac{S_Q(\dq)/S_Q(2\dq)-1}{4-S_Q(\dq)/S_Q(2\dq)}},
\quad \dq = \frac{2\pi}{L} \hat{\mathbf{x}},
\end{equation}
where $\mathbf{R}_\Delta$ is the position of trimer $\Delta$. Based on this 
standard Ornstein-Zernike form, we extract the length scale of the critical 
fluctuations in terms of the corresponding correlation function. 
We comment here that $S_Q$ encodes the $\langle \mathbf{S}^2_\Delta \mathbf{S}^2_{\Delta'} \rangle$ 
correlations, which would be an off-diagonal four-spin correlation function in 
the single-spin $S^z$ basis and thus comparatively difficult to access within SSE 
QMC. Employing the trimer basis and the methodology of Subsec.~\ref{sec:offdiag} 
is therefore beneficial not only in removing the sign problem of the FFTL but 
also in rendering the correlations crucial to describing its physics diagonal, 
and as a result easy to compute.

Figure~\ref{fig:correlen} shows scans of $\xi_Q$ across the phase diagram for 
the two cases $J_2 = 0$ and $J_2/J_1 = 1$. These plots look rather similar to 
the specific-heat data of Fig.~\ref{fig:ffts_specheat_scan}, with the notable 
exception that there is no trace of a second set of maxima to the left of the 
first-order line. From $\xi_Q$ we therefore identify a single characteristic 
line within the supercritical regime, which in essence continues the 
first-order line beyond $T_c$. Along this continuation line, fluctuations in 
the local trimer states proliferate, anticipating the critical point and then 
the first-order line that control the low-temperature physics. For comparison, 
we include in Fig.~\ref{fig:correlen} two other sets of characteristic 
quantities. One set is the lines of specific-heat maxima from 
Fig.~\ref{fig:ffts_specheat_scan}, although for further comparison we neglect 
the line of weak maxima to the left of $J_c$. The other is the line along which 
$\braket{l_\Delta} = 1$, a value we base on our estimate that $\braket{l_\Delta} 
\approx 1$ at the critical point. By analogy with the density of water at the 
critical point, we refer to this line as the “critical isochore.” 

In the supercritical regime, the lines of maxima from these three different 
thermodynamic quantities converge upon approaching the critical point for 
any value of $J_2$. Away from criticality, they begin to diverge, each 
apparently most sensitive to slightly different features of the non-divergent 
thermal fluctuations. It is notable that both the convergence near the critical 
point and the proximity of the characteristic lines further from criticality 
are best in the degenerate limit, $J_2/J_1 = 1$, when the first-order line is 
most “water-like.” Analogous observations regarding the loci of extrema of 
various response functions were also made in Ref.~\cite{Luo2014} with a view 
to determining the Widom line (strictly, the line of zero ordering field) in 
the critical phenomena of liquid-liquid transitions with differently slanted 
first-order lines. In this sense, the characteristic lines we have identified 
may all serve as quantum magnetic analogs of the Widom line, particularly for 
the doublet-degenerate FFTL.

\section{Conclusion}
\label{sec:conclusion}

We have presented a QMC approach to study highly frustrated quantum spin models 
in the trimer basis. The abstract directed-loop approach that we introduce 
requires no system-specific knowledge or bias. It generalizes easily to other 
computational bases, and thus should be of use for QMC simulations of many 
other classes of frustrated quantum spin system (for example those composed 
of four-spin clusters). Here we have applied our algorithm to the $S = 1/2$ 
Heisenberg antiferromagnet on the FFTL, where it performs efficient and 
sign-problem-free simulations up to large system sizes, allowing us to 
investigate the full finite-temperature physics of this model.

A key property of the FFTL model is the appearance of a first-order quantum 
phase transition between two AFM ground states that break the same symmetry. 
From the QMC simulations, we found that a line of first-order transitions 
emerges from this point at finite temperatures, and terminates at a thermal 
critical point in the 2D Ising universality class. We used a straightforward 
free-energy argument to understand the shape of the first-order line in terms 
of the low-lying excitations and the doublet entropy intrinsic to each trimer. 
What sets the FFTL apart from the quantum spin models studied to date in this 
context is the curvature of the first-order line induced by the doublet 
entropy contribution, which develops in the limit of degenerate doublets 
into a fully slanting line in the plane of temperature and coupling ratio. 

We demonstrated further that this increased slanting of the first-order line 
is associated with an evolution of the specific heat from the Ising limit, 
which features two symmetrical lines of maxima, to a single prominent line 
of maxima that appears as an extension of the first-order line into the 
supercritical regime. This effect can be explained within the Ising model
by a coordinate transformation in the vicinity of the critical point. We 
computed different characteristic thermodynamic quantities for the FFTL, 
notably the line of maxima in the correlation length and the critical 
isochore, to show that they align with the line of primary specific-heat 
maxima upon approaching the critical point from the supercritical regime, 
whereas the line of secondary maxima is not related to the critical-point 
physics. In this way we identified the quantum magnetic analog of the Widom 
line, serving as a continuation of the first-order line in the supercritical 
regime.

Our results show that the FFTL model is very valuable for exploring the 
physics of first-order (quantum) phase transitions and thermal criticality 
in the absence of symmetry-breaking  in frustrated quantum magnets. The 
model possesses a number of attributes that warrant further investigation. 
In particular, the first-order quantum transition separates two AFM ground 
states that break the same symmetry. In principle, this allows quantum 
fluctuations in an extended phase diagram to terminate the first-order 
transition at $T = 0$, leading to a quantum critical point. Beyond such a 
point, one may anticipate a smooth crossover between the $S = 1/2$ and $S
 = 3/2$ representations or an exotic intermediate phase that mixes the two. 
It is conceivable that this regime could be reached within the sign-free 
parameter subspace, for example by introducing a finite magnetic field or 
an anisotropy in the fully frustrated interactions, both of which would be 
accessible by the same abstract directed-loop QMC approach that we have 
developed here.

\section*{Acknowledgements}

We thank Fabien Alet for helpful discussions.
We acknowledge the support of the Deutsche Forschungsgemeinschaft (DFG, 
German Research Foundation) through Grant No.~WE/3649/4-2 of the program 
FOR 1807 and through project RTG 1995. We are grateful to the Swiss National 
Science Foundation, and to the European Research Council (ERC) for funding 
under the European Union Horizon 2020 research and innovation program (Grant 
No.~677061). We thank the IT Center at RWTH Aachen University and the JSC 
Jülich for access to computing time through JARA-HPC. 

\begin{appendix}
\section{Ground-state energy of the spin-3/2 Heisenberg model}
\label{sec:app_gs_energy}

\begin{figure}
	\centering
	\includegraphics{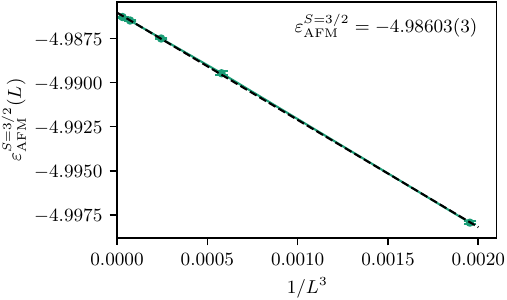}
	\caption{Ground-state energy per site, $\varepsilon^{S=3/2}_\mathrm{AFM} 
        (L)$ of the spin-3/2 Heisenberg model. The dashed line shows a fit to 
        Eq.~\eqref{eq:spin32exp}, which we use to extract the value in the 
        thermodynamic limit.}
	\label{fig:spin32groundstate} 
\end{figure}

The free-energy analysis of Subsec.~\ref{sec:ffts_transition} requires the 
ground-state energy per site of the spin-3/2 Heisenberg model on the square 
lattice. To compute this energy, we performed SSE QMC simulations of the model 
for several linear system sizes up to $L = 32$ and for inverse temperatures up 
to $\beta = 2L$. The AFM ground-state energy in the finite system converges to 
the thermodynamic limit with the form~\cite{Hasenfratz1993}
\begin{equation}
	\varepsilon_\text{AFM}^{S=3/2}(L) = \varepsilon_\text{AFM}^{S=3/2} + 
c/\beta{}L^2 + \dots,
\label{eq:spin32exp}
\end{equation}
where $c$ is a non-universal constant. We use this relation to extract 
the values $\varepsilon_\text{AFM}^{S=3/2} = - 4.98603(3)$ and $c = -12.1(1)$
(Fig.~\ref{fig:spin32groundstate}).

\end{appendix}

\nolinenumbers
\bibliography{references}

\end{document}